\pdfoutput=1

\documentclass[a4paper,10pt]{article}
\usepackage{amsmath,amssymb}
\usepackage[colorlinks]{hyperref}
\usepackage{eucal}
\scrollmode
\usepackage{graphicx}
\usepackage{epstopdf}
\usepackage{array, longtable}
\usepackage[english]{babel}
\title { A new model of the Central Engine of GRB and the Cosmic Jets\footnote{ Based on talk given on 15.06.2007 at The Advanced Workshop on Gravity, Astrophysics and Strings, GAS@BS07, 10-16.06.2007, Primorsko, Bulgaria.}}
\author {Plamen P. Fiziev\thanks{ E-mail:\,\,\,fiziev@phys.uni-sofia.bg},  Denitsa R. Staicova\thanks{ E-mail:\,\,\,denijane@gmail.com}
\\Department of Theoretical Physics, Sofia University ``St. Kliment Ohridski", \\ 5
James Bourchier Blvd., 1164 Sofia, Bulgaria }

\date{}
\begin{document}
\maketitle

\begin{abstract}
Despite all the already existing observational data, current models still cannot explain completely the excessive energy output
and the time variability of GRB. One of the reasons for this is the lack of a good model of the central engine of GRB.
A major problem in the proposed models with a black hole (BH) in the center is that they don't explain
the observed evidences of late time activity of the central engine.

In this paper we are starting the search for a possible model of that central engine as a rotating
compact body of still unknown nature. The formation of jets in the new model lies entirely
on the fundamental Teukolsky Master Equation.
We demonstrate that this general model can describe the formation of collimated GRB-jets of various forms.
Some preliminary results are presented.
\end{abstract}

\section{Introduction}
Gamma-Ray bursts have been mystifying scientists since their discovery in 1969.
Their spectral and temporal behavior raised many questions offering the unique opportunity
to confront our theories with reality. After all those years of discoveries and innovations,
one thing is clear -- GRB cannot be properly understood until there is a good model of the central engine.
In this paper, we are starting the study of a new model of the central engine of GRB
that can explain some of their most intriguing characteristics.
We do not wont to presuppose a large number of hypothetical and unconfirmed by observations
specific properties of the central engine. Instead, we prefer to focus our consideration
on the minimal number of natural assumptions,
which seem to be unavoidable in any model of central engine, and to investigate their possible
consequences.

\subsection{GRB in brief}
Before we present our model, we offer a summary of the properties of GRB that will be important in our studies.

Gamma-ray burst are explosions on cosmic distances (biggest measured redshift is $z > 6.3 $) 
that emit huge amounts of energy ( $\sim 10^{51}$ erg) in very short periods of time ($T_{90}\sim$ seconds) (\cite{Zhang}). 
They have two distinct phases -- prompt emission and afterglow -- produced by different mechanisms. 
An unexpected feature found by the mission SWIFT is the existence of flares (\cite{Zhang}, \cite{Burrows} and \cite{Falcone}) -- 
yet not completely explained peaks in the light curve superimposed over the continuing afterglow emission. 
The mechanisms of radiation that are believed to cause the burst are synchrotron emission and inverse compton scattering.

The widely accepted statement us that there exist 2 types of GRB -- short GRBs ( $T_{90} <2$ s) and long GRBs ($T_{90}>2$ s)
with some indications of a third type. The temporal separation is supposed to be not only empirical --
the two types may have different origin and different spectral characteristics.
Short GRBs are found in older galaxies, they have harder spectrum and they are thought
to be a product of binary mergers of compact objects -- black holes and/or neutron stars.
On the other hand, long GRBs are found in young, blue galaxies, with active star formation,
thus the idea they are a result of the collapse of massive stars(\cite{Zhang}).
This hypothesis was confirmed by the observation of GRB 060218 which evolved
to the supernova SN 2006aj (\cite{Mirabal} and \cite{Campana}).

Current development of the studies of GRB includes magneto-hydro-dynamical simulations of
jets propagating in stellar medium with different density or different structure of the jet,
testing the type of particles involved, the mechanisms of emission and also the type of object
that produces them. The observations of GRB continue to show interesting characteristics yet to
be studied and explained. Because of the dependency of the observations on the characteristics of the detectors,
mostly on their energy range, a new space missions are being prepared
-- GLAST\footnote{the Gamma-ray Large Area Space Telescope(GLAST ) was renamed to Fermi Gamma-ray Space Telescope
after the launch in 2008} among them -- that can observe GRB in higher energies which is important
for the good resolution of the usually hard prompt emission.

\subsection{Fireball model}
Fireball model is one of the most frequently used models in GRB physics \cite{Piran}.
In this model the central engine of GRB emits matter in series of shells with different Lorentz factors.
When the faster shells catch up with the slower ones, the resulting collision called internal shock,
produces the hard prompt emission. The deceleration of the shells due to the contact with the local medium
is believed to produce the softer afterglow. Major set-backs of this model are that it cannot explain the
late activity of the central engine of which the existence of flares is an evidence, nor the nature of those flares.

\section{A Toy Model of Central Engine and Generation of Relativistic Jets}

\subsection{General Description}

Our toy model is extremely simple and based on the least possible assumptions.
For a start, we have a Kerr black hole or some other rotating compact massive object,
both described by the Kerr metric (\cite{Kerr}): exactly -- in the first case,
and approximately -- in the second case.
In Boyer-Lindquist coordinates the metric is:
\begin{align}
 ds^2=(1-2Mr/\Sigma )dt^2+4 M a r\sin^2\left(\theta\right)/\Sigma dtd\phi-(\Sigma/\Delta)dr^2-\Sigma d\theta^2 -\sin^2(\theta)\left[r^2+a^2+2Ma^2r \sin(\theta)/\Sigma\right]d\phi^2
\end{align}
with $\Delta=r^2-2Mr+a^2, \Sigma=r^2+a^2 \cos^2\theta$.

For this metric, Teukolsky studied the linearized perturbations of the Einstein equations and found that the equations describing different perturbations generalize to \cite{Teukolsky2}:
\begin{equation}
\begin{split}
&L= \left[ \frac{(r^2+a^2)^2}{\Delta} -a^2\sin^2\theta\right]\frac{d^2}{d t^2}+\frac{4Mar}{\Delta}\frac{d^2}{d t d \phi}+\left[ \frac{a^2}{\Delta} - \frac{1}{\sin^2\theta}\right]\frac{d^2}{d \phi^2}-
\Delta^{-s}\frac{d}{d r}\left(\Delta^{s+1}\frac{d}{d r}\right)-\\ &\qquad \frac{1}{\sin(\theta)}\frac{d}{d \theta} \left(\sin\theta \frac{d}{d \theta}\right)-2s\left[\frac{a(r-M)}{\Delta}-\frac{i\cos\theta}{\sin^2\theta}\right]\frac{d}{d \phi}
-2s\left[\frac{M(r^2-a^2)}{\Delta}-r-ia\cos\theta\right]\frac{d}{d t}+\left(s^2\cot^2\theta-s\right).
\label{L}
\end{split}
\end{equation}

Following the procedure established by Teukolsky, we perform separation of the variables in the
Teukolsky equations using the following substitution:
\begin{equation}
 \Phi=e^{(\omega t+m\phi)i}S(\theta)R(r),
\label{phi}
\end{equation}
where m=...-2, -1, 0, 1, 2... is an integer, $\omega$ is the frequency,
and $S(\theta)$ and $R(r)$ are the angular and the radial part of the equation.
It is important to emphasize that the frequency $\omega$ is a complex number:
$\omega=\omega_R+i\omega_I$. Notice that we use different substitution for the time dependence: $ e^{i\omega t}$,  not the original one: $ e^{-i\omega t}$, that Teukolsky used.

One of the most important assumption we use is the stability condition $\omega_I>0$,
that ensures that the initial perturbation won't become infinite with time and it will damp instead.

Applying the operator (\ref{L}) L  on the function (\ref{phi}) $\Phi$ ,
Teukolsky found that the equations for $\theta$ and $r$ separate \cite{Teukolsky2} to an angular and a radial part.
The angular Teukolsky  Equation (TAE):
\begin{equation}
 \left[\left(1-u^2\right)S_{lm,u}\right]_{,u}+\left[(a\omega u)^2+2a\omega su+s+{}_sA_{lm}-\frac{(m+su)^2}{1-u^2}\right]S_{lm}=0.
\end{equation}
The Teukolsky radial equation  TRE:
\begin{equation}
\Delta R_{lm,rr}+2(s+1)(r-M)R_{lm,r}+V(r)R_{lm}=0,
\end{equation}
where $u=\cos(\theta)$, A is the constant of separation of the variables and the potential V(r) is:
\begin{multline*}
V(r)=-A-2\,is\omega\,r+ {\frac {{\omega}^{2}{r}^{2} \left( {r}^{2}+{a}^{2}
 \right) +2\,is \left( ma \left( r-M \right) +\omega\,M \left( {r}^{2}
-{a}^{2} \right)  \right) +2\,\omega\,Mr \left( \omega\,{a}^{2}+2\,am
 \right) +{m}^{2}{a}^{2}}{\Delta}}.
\end{multline*}

Here, s=-2, -1, -1/2, 0, 1/2, 1, 2 is the spin weight, related to the value of the spin $|s|$ of the field.
It is remarkable that all fields can be described with one set of equations.
Also, it is important that ${}_sA_{lm}$ and $\omega$ are independent parameters.

To solve that equation, we use the standard notations: $r_{_+\!}\!=\!M+\sqrt{M^2\!-\!a^2}$ is
the event horizon and $r_{_-}\!=\!M-\sqrt{M^2\!-\!a^2}$ is the Cauchy horizon.
It's easy to see that there is a symmetry between $r_{_+}$ and $r_{_-}$ in the TRE
and that $r_{_+}$ and $r_{_-}$ are regular singular points of the TRE,
while $r=\infty$ is an irregular singular point.

Using the software package Maple to solve that linear differential equation we acquire two
independent exact solutions of the radial Teukolsky equation in the outer domain ($r>r_{_+}$) \cite{Fiziev06a}:
\begin{align}
&R_1 \left( r \right) ={\it C_1}\,{{ e}^{-i\omega\,r}} \left( r-{\it
r_{_+}} \right) ^{{-i\frac { \omega\,({a}^{2} + {{\it r_{_+}}}^{
2})+am }{{\it r_{_+}}-{\it r_{_-}}}}} \left( r-{\it r_{_-}} \right) ^{{
-i\frac {\omega\,({a}^{2}+{{\it r_{_-}}}^{2})+am}{
{\it r_{_+}}-{\it r_{_-}}}}+1}
{\it HeunC} \left( \alpha,\beta,\gamma,\delta ,\eta,{\frac {r-{\it r_{_+}}}{-{\it r_{_+}}+{\it r_{_-}}}} \right)
\end{align}
and
\begin{align}
 &R_2 \left( r \right) ={
\it C_2}\,{{ e}^{-i\omega\,r}} \left( r-{\it r_{_+}} \right) ^{{i\frac {\omega\,({a
}^{2}+{{\it r_{_+}}}^{2})+am}{{\it r_{_+}}-{\it r_{_-}
}}+1}} \left( r-{\it r_{_-}} \right) ^{{
-i\frac{\omega\,({a}^{2}+{{\it r_{_-}}}^{2})+am}{
{\it r_{_+}}-{\it r_{_-}}}+1}}
{\it HeunC} \left( \alpha,
-\beta,\gamma,\delta ,\eta,{
\frac {r-{\it r_{_+}}}{-{\it r_{_+}}+{\it r_{_-}}}} \right)
\end{align}
where HeunC is the confluent Heun function (see \cite{Heun} and \cite{Heun2}) and its parameters are  :
\begin{align*}
\alpha &=2\,i \left( {\it r_{_+}}-{\it r_{_-}}
 \right) \omega, &
\beta &={-\frac {2\,i(\omega\,({a}^{2}+{{\it
r_{_+}}}^{2})+am)}{{\it r_{_+}}-{\it r_{_-}}}}-1,\\
\gamma &={-\frac {2\,i(\omega\,({a}^{2}+{{\it
r_{_-}}}^{2})+am)}{{\it r_{_+}}-{\it r_{_-}}}}+1, &
\delta &=-2\!\left({\it r_{_+}}-{\it r_{_-}} \right)\!\omega\!\left( i+
 \left( {\it r_{_-}}+{\it r_{_+}} \right) \omega \right).
\end{align*}

\begin{align*}
\eta =\!\frac{1}{2}\frac{1}{{
\left( {\it r_{_+}}-{\it r_{_-}} \right) ^{2}}}\Big[ 4{\omega}^{2}{{\it r_{_+}}}^{4}+ \left(4i
\omega -8{\omega}^{2}{\it r_{_-}}\right) {{\it r_{_+}}}^{3}&+ \left( 1-4a\omega\,m-2{\omega}^{2}{
a}^{2}-2A \right)  \left( {{\it r_{_+}}}^{2}+{{\it r_{_-}}}^{2} \right) + \\
 &\left(4\,i\omega\,{\it r_{_-}} -8i\omega\,{\it r_{_+}}+4A-4{\omega}^
{2}{a}^{2}-2 \right) {\it r_{_-}}\,{\it r_{_+}}-4{a}^{2} \left( m+\omega\,a
 \right) ^{2} \Big].&
\end{align*}

The angular equation has regular solutions studied by Teukolsky and Press(\cite{Teukolsky3}.
Beside those solutions, one can find polynomial solutions of TAE.
As far as we know these polynomial solutions haven't been studied, being singular.
Explicitly, two solutions of TAE in terms of confluent Heun functions are:

\begin{align*}
S_{{\pm},s,m}^{(-1)}(\theta)={{\rm e}^{\pm \Omega\,\cos\theta }}& \left(\cos\left(\theta/2\right)\right)^{\mid s-m \mid}
  \left(  \sin \left(\theta/2\right)\right) ^{- \mid s+m \mid}\times\\&
  {\it HeunC} \left( \pm 4\,\Omega,\mid\! s\!-\!m\! \mid,\mid\! s\!+\!m\! \mid,-4\,\Omega\,s,\frac{{m}^{2}-{s}^{2}}{2}+2\,\Omega\,
s-{\Omega}^{2}-A-s,  \cos^2 \frac{\theta}{2}    \right), \\
\end{align*}
and
\begin{align*}
S_{{\pm},s,m}^{(1)}(\theta)={{\rm e}^{\pm \Omega\,\cos \left( \theta \right) }}& \left(\cos \left( \theta/2\right) \right)^{\mid s-m \mid} \left(\sin \left( \theta/2 \right)\right)
 ^{- \mid s+m \mid}\times\\
&{\it HeunC}  \left( \mp 4\,\Omega,\mid\! s\!+\!m \!\mid,\mid\! s\!-\!m\! \mid,-4\,\Omega\,s,\frac{{m}^{2}-{s}^{2}}{2}-2\,\Omega\,s
-{\Omega}^{2}-A-s,  \sin^2 \frac{\theta}{2}
  \right),
\end{align*}
where $\Omega=a\omega$.
In general, these are two linearly independent solutions of the TAE.
They arise from the properties of the Heun functions.
A special attention is required in the case $s \in \mathbb{Z}$ -- then $s\!-\!m \in \mathbb{Z}$
and the second linearly independent solution is not in the above form,
but includes an integral of the confluent Heun function.

\subsection{Numerical Results}
Once equipped with the solutions, we proceed with numerical search for the explicit value of the frequency $\omega$.
A major technical problem in the use of the regular solutions of the angular equation is that we are trying
to solve connected spectral problem with two complex solutions.
It has been solved for the first time by Press \& Teukolsky (\cite{Teukolsky3})
and after that developed trough the mechanism of continued fractions by Leaver \cite{Leaver}.

In our work, the case is being simplified drastically by the consideration
of polynomial solutions, i.e., by the assumption that the confluent Heun functions
in the angular equation are polynomials.
From here, we manage to obtain a simple analytic relation $A(\omega)$.
Using the explicit form of $A(\omega)$ we can plot the angular part of the solution
(\ref{phi}) for certain values of $\omega$.

\begin{figure}[!htb]
 \centering
 \includegraphics[width=400px,height=230px,bb=0 0 782 448]{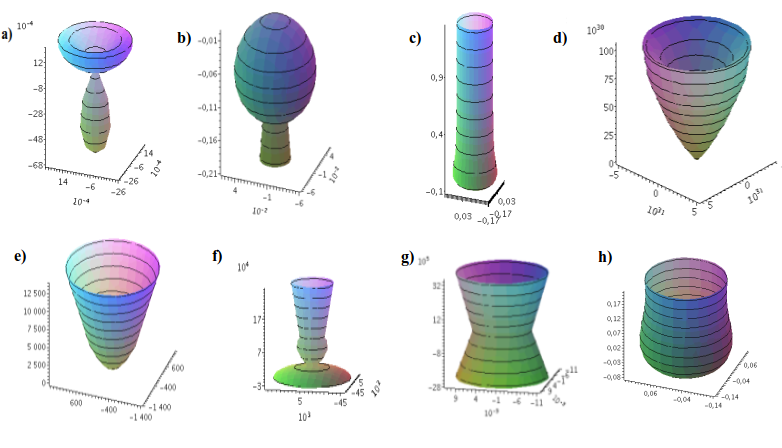}
\caption{Few examples of jets that we obtained from the solutions of the angular equation.}
 \label{fig:jets}
\end{figure}

An important feature of the singular polynomial solutions is that they provide
a natural explanation for the existence of jets.
Examples of such jets in the case s=-1 can be seen on figure \ref{fig:jets}.
In that figure, one can see both limited and unlimited solutions,
although the second one can be constrained adding a form factor with known value.
Also, we see that different form of the jets arise for different m.
Note that the case $|s|=1$, presented in this talk, corresponds to perturbations,
which describe electromagnetic waves.
An analogous treatment of the case $|s|=2$, which is more complicated and corresponds
to gravitational waves, is presented in \cite{Fiziev06b} and \cite{PFDS}.

The general solution we use to obtain those plots for $|s|=1$ is:
\begin{align*}&S(\theta)={\left( \cos \theta\!+\!1 \right)^{\frac{1+m}{2}} \left( 1\!-\!\cos \theta \right) ^{\frac{m-1}{2}} {\rm e}^{\Omega\,\cos \theta  }}{\it HeunC} \left( 4\,
\Omega,m\!+\!1,m\!-\!1,4\,\Omega,\frac{1}{2}{m}^{2}\!-\!A-2\,\Omega\!-\!{\Omega}^{2}\!+\!\frac{1}{2},\cos^2 {\frac\theta 2}  \right). \end{align*}

Using the properties of the confluent Heun function and imposing the polynomial conditions
on the solutions of the angular equation, we obtain the following polynomials, summarized in Table 1.
\begin{table}[!htb]
\begin{tabular}{|m{30px} | m{350px} | m{60px} |}
 \hline Figure  & Formula  & Parameters \\ \hline
& &\\
  a), b) & {\begin{align*}\frac {{{\rm e}^{-\Omega\cos\theta }} \left( 1+ \left( -2\,\Omega \mp 2\,\sqrt {{\Omega}^{2}-\Omega} \right)  \cos^2\!\frac{\theta}{2} \right) }{  \sin^2\!\frac{\theta}{2}}\end{align*}}
 & $L\!=\!-m=\!1$ \\
& &\\
c) & ${{{\rm e}^{-\Omega\cos\theta }} \cot \frac{\theta}{2}}$ & $m\!=\!L=\!0$ \\ & & \\
d)& {\begin{align*}
&{{\rm e}^{-\Omega\cos\theta }} \left(\cos{\frac{\theta}{2}}\right)^{-(m+1)}  \left( \sin {\frac{\theta}{2}}
\right) ^{m-1} \left( 1+ \frac { \left( 2 \,\Omega+2\,\sqrt {\Omega\, \left( \Omega+m \right) } \right) \sin^2 \frac{\theta}{2}}{m} \right)  \end{align*}
}
& $L\!=\!1$, $m\!=\!6$\\
 e), f) &
{\begin{align*}
 {{{\rm e}^{-\Omega\cos\theta }} \left( 1+ {\tan^{2} \frac{\theta}{2}}
 \left( 2\,\Omega \mp 2\,\sqrt {{\Omega}^{2}+\Omega} \right)   \right) } \end{align*}
}  & $L\!=m\!=1$\\
g) & ${{{\rm e}^{-\Omega\cos\theta }} \tan \frac{\theta}{2}}$ & $m\!=\!L=\!0$ \\ & & \\
h) & {\begin{multline*}{{\rm e}^{-\Omega\cos\theta }} \left(\cos  {\frac{\theta}{2}}\right) ^{-m-1} \left(\sin { \frac{\theta}{2}}\right)^{m-1} \left( 1+{\frac { \left( 2
\,\Omega+2\,\sqrt {\Omega\, \left( \Omega+m \right) } \right)  \left(
\cos^2 \frac{\theta}{2} \right)}{m}} \right)
\end{multline*}}  &$L\!=\!-m\!=\!1$ \\ \hline
\end{tabular}
\caption{A table of the functions used to plot Figure 3}
\end{table}

The radial equation is much harder to deal with. Using $A(\omega)$ and the right boundary conditions for a black hole, we want to look for zeroes of the resulting functions. This task is hard considering the profile of the confluent Heun functions.

We use the distributions of the eigenvalues in the complex plane for the singular case: $s=-1,\, m=1$ for certain  $A(\omega)$
\big(explicitly $A_{s=-1, m}(\omega)\!=\!-{\Omega}^{2}\!-\!2\,\Omega\,m\!\pm\!2\,\sqrt {{\Omega}^{2}\!+\!\Omega\,m}$ \big)
to find points resembling to zeroes of the function.
After that we need an algorithm that can prove they are exactly a zeros of the function and not just a minimum.

Such new algorithms for searching for the complex roots of a complex function are being developed by the team. The results will be reported in the following paper.

Another problem is that the numerical calculations are additionally burdened by the CPU time required for making a complex plot of the solutions by the package Maple.

\section{Conclusion}
Although in very preliminary stage, our model of central engine seems to be able to produce qualitatively
some of the basic features observed in GRB.
The formation of relativistic jets is supposed to be caused by the rotation of a compact central body
of any nature.
At least in a good approximation its exterior gravitational field can be described by Kerr solution.
The Teukolsky Master Equation (with appropriate boundary conditions) is fundamental enough
to account for both types of GRB with their maybe different origin.
The essential assumption we used is that the imaginary part of the frequency should be positive.
This provides stability of the solutions.
Different objects can be described by different boundary conditions,
though in our case, we used the standard black hole boundary conditions for the radial Teukolsky equation:
on the horizon we have only entering waves and on infinity we have only outgoing waves.
In contrast, we used a novel boundary conditions for the angular Teukolsky equation and this enables us
to describe mathematically the collimated jets of different forms.
Our preliminary results show the potential of this mathematical model for description of
the central engine as a rotating relativistic compact object of any nature.
The development of our study will be published in the papers to follow.

\vskip .5truecm
{\bf Acknowledgements}
\vskip .3truecm
This paper made use of data supplied by the UK Swift Science Data Centre at the University of Leicester.

Our work is supported by the Foundation "Theoretical and
Computational Physics and Astrophysics"
and by Bulgarian National Scientific Fund under contracts DO-1-872, DO-1-895 and DO-02-136.

\end{document}